# Screening and understanding Li adsorption on 2-dimensional metallic materials by learning physics

Sheng Gong[1,*], Shuo Wang[2,*], Taishan Zhu[1,*], Xi Chen[1] and Jeffrey C. Grossman[1,†]

[1]Department of Materials Science and Engineering, Massachusetts Institute of Technology, MA 02139, USA

[2]Department of Materials Science and Engineering, University of Maryland, MD 20742, USA

# ABSTRACT

Two-dimensional (2D) materials have received considerable attention as possible electrodes in Li-ion batteries (LIBs), although a deeper understanding of the Li adsorption behavior as well as broad screening of the materials space is still needed. In this work, we build a high-throughput screening scheme that incorporates a learned interaction. First, density functional theory and graph convolution networks are utilized to calculate minimum Li adsorption energies for a small set of 2D metallic materials. The data is then used to find a dependence of the minimum Li adsorption energies on the sum of ionization potential, work function of the 2D metal, and coupling energy between $Li^+$ and substrate. Our results show that variances of elemental properties and density are the most correlated features with coupling. To illustrate the applicability of this approach, the model is employed to show that some fluorides and chromium oxides are potential high-voltage materials with adsorption energies < -7 eV, and the found physics is used as the design principle to enhance the Li adsorption ability of graphene. This physics-driven approach shows higher accuracy and transferability compared with purely data-driven models.

## INTRODUCTION

Lithium-ion batteries (LIBs) have been studied for decades with continued interest in the search for new battery materials with higher energy density, higher stability and faster kinetics. Two-dimensional (2D) materials have generated excitement for application in LIBs[1] because of their distinct structural characteristics such as enlarged interlayer spacing and high aspect ratio which can help alleviate structural instabilities, accelerate faster ion diffusion and enhance Li adsorption[2, 3, 4]. Recently, 2D materials have been used as electrodes[5, 6] including as composite materials such as in the form of mixed[7, 8, 9], wrapped[7, 8, 10], or encapsulated[7, 8, 11, 12] composites and layered heterostructures[7, 13, 14], showing a broad application of 2D materials in the field of LIBs. Despite this promise, the interactions between Li and 2D materials are not well understood due to the lack of both experimental and computational data[15, 16].

Currently, there are three main 2D materials databases with atomic and electronic structures from *ab* initio methods, C2DB[17], Materials Cloud[18] and 2D Materials Encyclopedia[19], and Jain *et al.*[20] summarized the three databases into one comprehensive database with 7,736 2D materials. One of the most important properties to be computed in order to assess the use of one of these materials as an electrode in LIB is the Li adsorption energy[21], which requires *ab* initio methods such as density functional theory (DFT) to predict with sufficient accuracy across a large range of materials. Yet, it is challenging to screen over a large number of these materials with DFT since for each 2D material there are multiple possible adsorption sites, and different from the case of H adsorption where only "high-symmetry-sites" are screened[22, 23], Li is known to be adsorbed on sites far away from high-symmetry points[11, 24], which leads to a massive set of possible systems.

With the rapid development of machine learning (ML) techniques in the field of materials science, ML models are now frequently used in high-throughput screening[25, 26, 27, 28], where they are typically first trained on a small subset of the given materials space then directly applied to screen the larger materials space of interest. In the process, physical insight can be extracted as shown schematically in Figure 1a. Since Li adsorption is site-dependent, here the most intuitive way to screen the minimum Li adsorption energy is to build machine learning potentials for all the materials and find the minimum adsorption energy from the potential-energy surface. However, it is still challenging to construct a universal potential that can generalize well for thousands of materials[29, 30], as will be shown later. Learning the minimum adsorption energy directly from the structure of the substrate is also not feasible, as it is too computationally demanding to generate the massive data of minimum Li adsorption energies, which limits the capacity of machine learning models[31], and the low transferability of current machine learning models also makes the prediction on unseen data unreliable[32, 33, 34]. Therefore, it is important to explore approaches that go beyond conventional machine learning-based high-throughput screening that rely on purely data-driven machine models. Here we incorporate physical insights related to Li binding behaviors in order to simplify the learning problem.

For Li insertion into bulk materials, one approach is to estimate the energetics as a charge transfer process between different energy levels[35, 36, 37], while leaving out other interactions, especially electrostatic due to their high complexity. Li adsorption on surfaces, including 2D materials, is considered to be a chemisorption process[38, 39], in which the Li 2s orbital hybridizes with surface valence levels, and the energy change depends on the degree of filling of the antibonding orbital. This understanding explains experimental observations well[40] and provides good qualitative trends. However, it is more challenging to estimate energetics quantitatively, especially when screening across different materials. In a recent work by Liu et al.[41], Li adsorption is modeled via a two-step process: a Li atom first ionizes to a $Li^+$ and an electron, and then the Li$^+$ couples with the charged 2D materials electrostatically. The system can be then approximated as an image-charge coupling where Li$^+$ is a +1 point charge and the 2D material is represented as a continuous conducting plate, giving the adsorption energy estimate as:

$$E_{ads} = IP - 14.38/(2*h) \quad (1),$$

where E$_{ads}$ is minimum Li adsorption energy in eV, IP is the ionization potential of Li in eV and h is adsorption height in Å. Although this estimation explains trends between different alkali ions well[41], the electronic structure of the substrate material is not accounted, which limits use across different materials.

Because it is hard to screen the minimum Li adsorption energy by either DFT calculations or conventional purely-data driven machine learning models, and because we know that the minimum Li adsorption energy is related to the process of ionization, charge transfer, and coupling, in this work we incorporate the process of learning physics into the screening circle (shown schematically in Fig. 1b). The aim of this approach is to simplify the screening problem by decoupling the target property into simpler properties for higher efficiency, accuracy and transferability. First, we build a subset with 5% of the 2D materials that are metallic (we restrict the present study to metallic materials since high electrical conductivity is desired in electrodes and they share different Li binding behaviors compared with semiconductors as will be discussed later). This subset of materials is then used to train graph convolutional network (GCN)[30, 42, 43, 44] potentials to learn the adsorption energies with actively sampled adsorption sites. With the well-trained potential we then calculate the minimum Li adsorption energy for each material in the subset, and we use the data of the subset to fit and propose the physics of Li adsorption, and then attempt to further examine and understand the physics based on a combination of human knowledge and machine learning. We find that the Li adsorption energy is the sum of three terms: ionization potential of Li, fermi level (work function) of the 2D metal, and coupling energy between Li$^+$ and the negatively charged substrate, . Finally, we use these correlations to screen the remaining 2D metals and find that fluorides and chromium oxides could be promising high-voltage materials, and we use the found physics as the design principle to enhance Li adsorption ability of graphene. We show that the physics-based model has higher accuracy and transferability than the purely data-driven machine learning models, and beyond screening, key understanding derived from the physical model also helps to evaluate existing theories and raise new questions for further study.

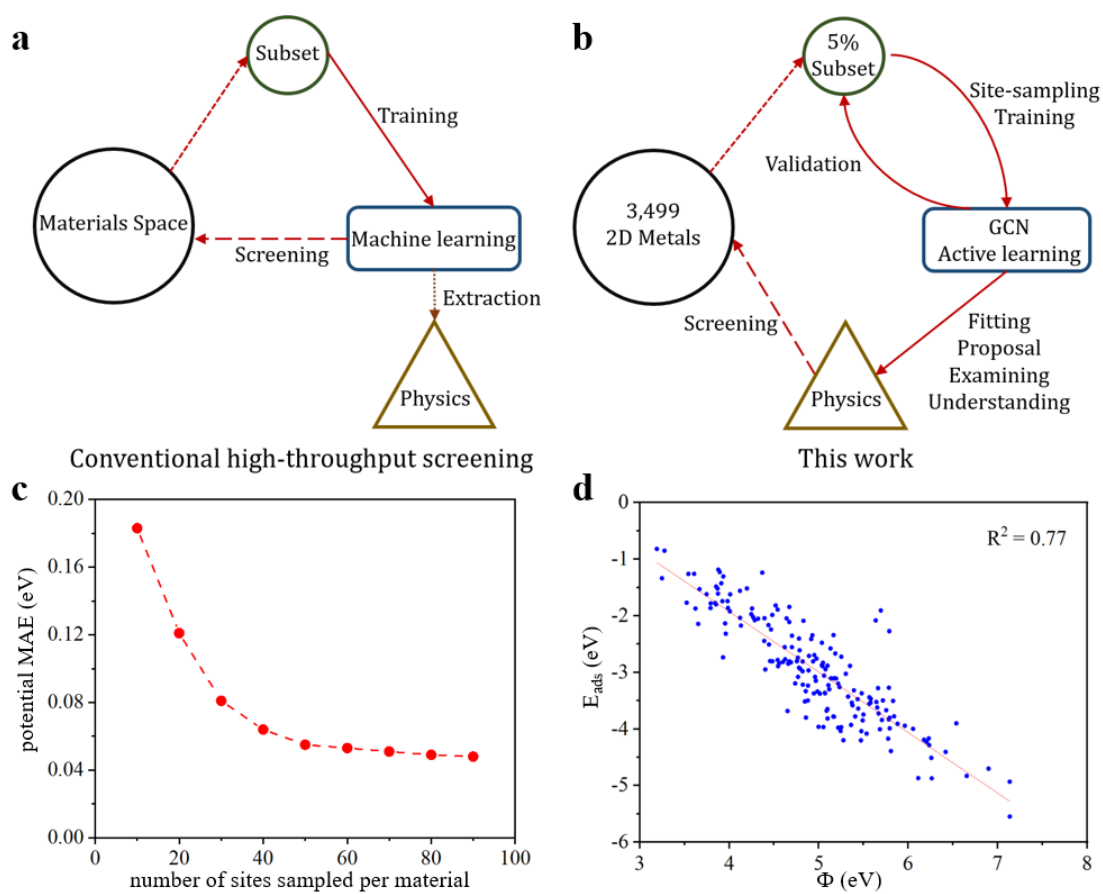

**Figure 1.** **a** and **b** Schematics of conventional machine learning-based high-throughput screening workflow and the approach adopted in this work, respectively. **c** Mean Average Error (MAE, in eV) of the potentials for test sets versus number of adsorption sites sampled per material. **d** Minimum Li adsorption energy ($E_{ads}$, in eV) versus work function ($\Phi$, in eV).

## RESULTS AND DISCUSSIONS

**Minimum Li adsorption energies from GCN and DFT.** In this work, we choose graph convolutional networks (GCN) as the machine learning architecture for learning Li site energies at different adsorption sites, because it has been shown to encode atomic and geometric information with high transferability[34, 43, 45], and has been utilized as a model form of interatomic potentials[30, 42]. In order to efficiently learn Li adsorption energies at different sites, we iteratively sample sites from each material with site energies calculated by DFT, and then train GCN on all the calculated energies. For validation, after each training we sample a new set of site energies, and these energies are added to the training data for the next generation GCN potential, until the validation error converges. The convergence is illustrated in Figure 1c, from which one can see that the error of the predicted site energies converges at around 0.05 eV. As a comparison, a GCN potential trained on all 3,499 2D metals, with the same number of total site energies as the training set (around 15,000), gives an error of ~0.5 eV, or 10 times higher than the error for the subset, which illustrates the difficulty of building a universal machine learning potential that generalizes well for the large materials space. Here we use the trained GCN potential to determine the minimum adsorption site for each material in the subset, and DFT optimization is followed for calculation of accurate minimum adsorption energies.

**Linear relation between Li adsorption energy and work function.** From the aforementioned understanding of charge transfer between different energy levels, we expect that the minimum Li adsorption energy should have a strong correlation with the position of lowest unoccupied band, or work function for zero-gap 2D materials. As shown in Figure 1d, the minimum Li adsorption energy has a strong linear relation ($R^2 = 0.77$) with work function:

$$E_{ads} = -1.07\Phi + 2.35 \ (2),$$

where Φ is the work function of the 2D metal in eV. In order to understand this linear relation, we propose a three-step adsorption mechanism based on the understanding described earlier of charge transfer and ionization-coupling mechanism, as shown in Figure 2a. When approaching a 2D metal, a Li atom first ionizes to a $Li^+$ and an electron, which costs the energy equal to the ionization potential of Li, then the electron transfers to the 2D metal and releases energy equal to the work function, and finally the $Li^+$ couples with the negatively charged 2D metal with energy change of the coupling energy. According to this description, the minimum Li adsorption energy should have three terms:

$$E_{ads} = IP - Φ + E_{cp} \ (3),$$

where $E_{cp}$ is the coupling energy between $Li^+$ and the negatively charged substrate in eV.

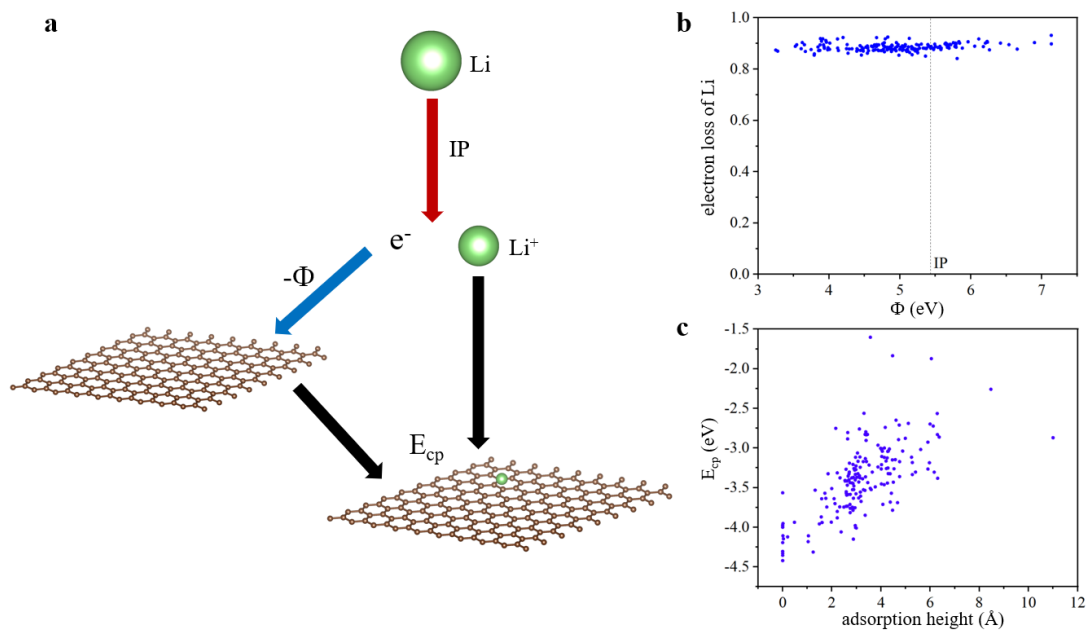

**Figure 2. a** Illustration of the proposed three-step adsorption mechanism. Here Φ is the work function, IP is the ionization potential of Li and $E_{cp}$ is coupling energy between $Li^+$ and the negatively charged substrate. **b** Electron loss of Li versus work function (Φ, in eV). The vertical line in the middle denotes

the position of Li ionization potential. **c** Coupling energy ($E_{cp}$, in eV) versus adsorption height (from $Li^+$ to the central plane of the 2D material, in Å).

Although equation (3) looks very similar to equation (2), there is one "counterintuitive" behavior in our proposed mechanism from the understanding of charge transfer. In Figure 2a, we assume that the electron always transfers from the Li atom to the 2D material, and Figure 2b shows that, from bader charge analysis[46], in all cases Li indeed transfers ~0.9 electron to the 2D metals. This cannot be described by a picture wherein charge transfer results from the difference in energy levels, since for 2D metals with work function lower than the ionization potential of Li, their fermi levels are higher than the Li 2s level, and electrons would transfer from substrates to the Li atom if electrons always flow from high energy levels to low ones. Based on the observed electron transfer direction, charge transfer cannot simply occur from one isolated level to another, and according to chemisorption theory, the DFT-computed electron transfer from Li to the substrate arises because of a broadening and shift of the Li 2s orbital as it approaches the substrate[38, 47]. Although not explicitly expressed in the proposed formula, these effects are included in our mechanism to support the proposed behavior.

Different from the proposed formula (equation (3)), the slope of the line fitted from the subset (equation (2)) is not -1, but rather -1.07. This deviation is likely due to bias of the data, because although the bonding between Li and 2D materials is not purely ionic[39], there is no correlation ($R^2 = 0.04$) between work function and charge transfer (Figure 2b), and therefore there is no relation between the ratio of ionic/covalent bonding and work function. When we set the slope to be -1 and fit the intercept from the subset, we have a new fitted line:

$$E_{ads} = -\Phi + 2.01 \ (4).$$

Comparing the numerical accuracy between equation (2) and equation (4), we find that the mean average error (MAE) and $R^2$ score between equation (2) and GCN/DFT-based data are 0.339 eV and 0.765, respectively, while that for equation (4) are 0.343 eV and 0.761, respectively, showing that switching from equation (2) to equation (4) doesn't have high impact on numerical accuracy. Based on this analysis and comparison, we choose equation (4) as the determined linear relation for the following parts.

**Prediction of the coupling energy.** As shown in equation (1), if $Li^+$ is considered a +1 point charge and 2D materials are continuous conducting plates, then the coupling energy ($E_{cp}$) is:

$$E_{cp} = -14.38/(2*h) \quad (5).$$

As shown by the results in Figure 2c, the trend of lower adsorption height, stronger coupling holds reasonably well at least qualitatively. Note, however that for cases with 0-height adsorption, which occurs for porous 2D materials[11], it is no longer reasonable to approximate 2D materials as continuous plates. The image-charge coupling also implies that the interaction between Li and substrates is purely ionic, which is not the case according to the charge transfer (Figure 2b) and experimental observations based on NMR measurements[39].

In order to improve our understanding of this coupling and predict $E_{cp}$ without calculating adsorption height by DFT, we build a random forest model to learn $E_{cp}$ from compositional and structural features as:

$$E_{cp} = RF(features) \quad (6),$$

where RF represents the random forest model, and features denote the compositional and structural features listed in Supplementary Information. As a result, we have a new equation for the adsorption energy:

$$E_{ads} = IP - \Phi + E_{cp} = IP - \Phi + RF(features) \quad (7).$$

As shown in Table 1, using a random forest model lowers the mean average error (MAE) of the linear relation with fixed intercept (equation (4)) by one-third where the $E_{ads}$ is solely dependent on $\Phi$ and as a result $E_{cp}$ is seen as a constant value. Based on the data shown in Table 1, we can also see that equation (7) has higher accuracy than directly learning $E_{ads}$ by random forest since the problem of predicting $E_{ads}$ is decoupled into two separate problems, where DFT is used to accurately and efficiently calculate the work function $\Phi$, which can be derived from a single-point calculation for each material. If $E_{ads}$ is learned directly, then we essentially learn $E_{cp}$ and $\Phi$ at the same time (as shown in the feature importance in Figure S1), which makes the learning problem harder. This helps to understand why the MAE for learning $E_{ads}$ is larger than that for learning $E_{cp}$ and $\Phi$ as shown in Table 1.

Table 1. Comparison of mean average error (MAE) of predictions from different models from leave-one-out cross-validation. Here "RF(features)" means that the quantity is from a random forest model with selected features described in the Supplementary Information, "$\bar{E}_{cp}$" denotes the mean value of $E_{cp}$, "IP" is the ionization potential of Li and the "$\Phi$" terms in right two cases are work functions from DFT. For the right two cases, the two models in each case share the same MAE according to equation (7).

| Models | $\Phi$ = RF(features) | $E_{ads}$ = RF(features) | $E_{cp}$ = RF(features) $E_{ads}$ = IP − $\Phi$ + RF(features) | $E_{cp} = \bar{E}_{cp}$ $E_{ads}$ = − $\Phi$ + 2.01 |
|---|---|---|---|---|
| MAE (eV) | 0.268 | 0.345 | 0.219 | 0.343 |

In order to further understand $E_{cp}$, we show the $R^2$ scores between compositional and structural features and $E_{cp}$ in Figure 3a, from which we can see that deviation and range of elemental properties and density-related features are those with the strongest correlation with $E_{cp}$. The latter is more straightforward to understand than the former. For density, as shown in Figure 3c, the less densely packed, the stronger the correlation since for less densely packed 2D sheets, $Li^+$ is more likely to be adsorbed into the pores with lower adsorption height. Variances of elemental properties are known to be important for thermal and mechanical properties[28], however, they are much less well understood in terms of correlation with Li adsorption. As shown in Figure 3b, more similar components, the stronger coupling. One explanation for this trend is that, more similar components, the bonds in the 2D materials become less polarized and more covalent because of more similar atoms at ends of bonds, and because more covalently bound materials tend to contain poorly packed structures with corner-shared or isolated polyhedral[35], it is more likely that those materials become porous with pores for $Li^+$ to be adsorbed in, as shown in Figure S2.

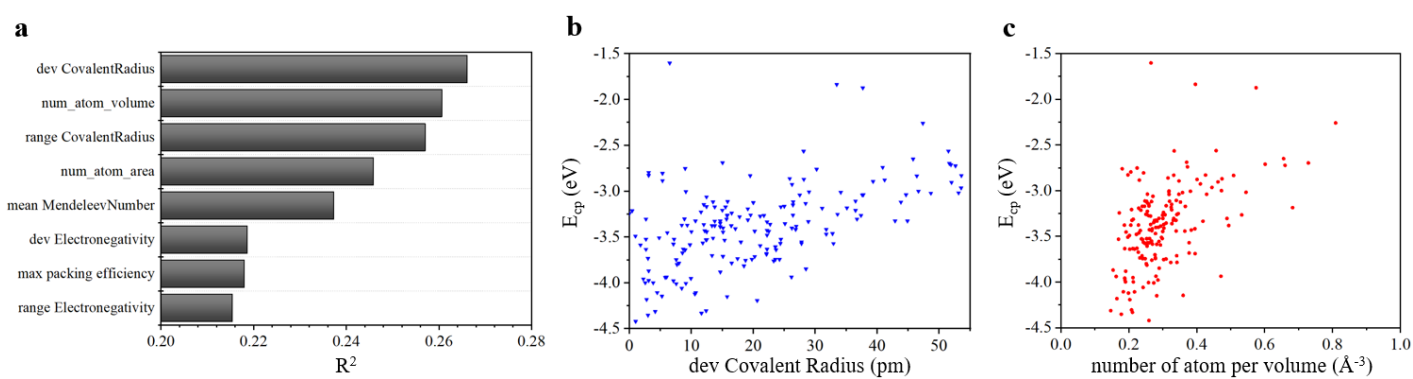

**Figure 3. a** $R^2$ scores between compositional and structural features and $E_{cp}$. Here only features with $R^2$ larger than 0.2 are plotted. **b** $E_{cp}$ versus standard deviation of covalent radius of elements in the 2D materials (in pm). **c** $E_{cp}$ versus number of atoms per volume (area*thickness, in Å$^{-3}$).

**Transferability and screening high-voltage materials.** After building up the linear dependence of the work function and random forest model for coupling, we use the models to screen all the 2D metallic materials to test transferability and search for promising battery materials. As an example, we select materials with the lowest minimum Li adsorption energies from our models as potential high-voltage materials, and then use DFT to manually calculate minimum Li adsorption energies, with results listed in Table 2. We find that some fluorides and chromium oxides are promising high-voltage materials with minimum Li adsorption energies lower than -7 eV, which exceeds the previous records for 2D materials of ~-5 eV[15]. Although the Li adsorption ability of the mentioned structures are newly examined here, the systems of fluorides and chromium oxides have already been considered as strong Li-binding substance because of the high electronegativity of F[35, 37] and the variability of oxidation state of Cr[48]. Since lower adsorption energy (higher binding strength) corresponds to higher voltage and higher energy density of electrode, and the voltages of most of the current cathode materials are lower than 5V[49], the found materials with ultrahigh Li-binding ability show great potential for improving energy density of LIBs[50]. As shown in Table 2, the predictions of ultrahigh binding strength are confirmed by the DFT calculations, which highlights the ability of our physics-driven models to extrapolate, since the lowest adsorption energy in the training set is around $-6$ eV. As a comparison, the purely data-driven model, $E_{ads} = $ RF(features), fails to identify the found high-voltage materials, even though it has similar performance on the training set compared with equation (4), showing the poor transferability of the purely data-driven model.

Table 2. Prediction of minimum Li adsorption energy (in eV) on 2D metallic materials from different models. Materials with the top 5 lowest minimum Li adsorption energies from equation (7) are included, and the numbers after the two $N_6F_{30}$ are their IDs in the database from Jain *et al.*[20], respectively.

|  | $E_{ads} = -\ \Phi + 2.01$ | $E_{ads} = IP - \Phi + RF(features)$ | $E_{ads} = RF(features)$ | DFT |
|---|---|---|---|---|
| $Bi_2F_2$ | -7.29 | -7.21 | -4.64 | -6.82 |
| $N_6F_{30}$ (2431) | -7.14 | -7.19 | -4.97 | -7.48 |
| $N_6F_{30}$ (2783) | -6.87 | -7.13 | -3.93 | -7.35 |
| $Cr_9O_{27}$ | -7.24 | -7.13 | -3.93 | -7.10 |
| $Cr_4P_8O_{32}$ | -7.25 | -7.12 | -4.53 | -8.04 |

**Enhancing Li adsorption ability of graphene.** In addition to screening new materials, the found physics can also serve as design principle to modify Li adsorption ability of existing materials. As an example, here we try to enhance the Li adsorption ability of graphene, which is known to have superior mechanical strength and electrical conductivity[21] while suffer from weak Li adsorption[51]. Because of the found linear relation between work function and Li adsorption energy, here we try to rise the work function of graphene by two types of p-type doping, B-doping and F-functionalization. For each modification, we construct three defect concentrations, and Li is adsorbed on the same site for all the six defect graphene structures, as shown in Figure S3. In Figure 4, the relation between the work function change and Li adsorption energy change is plotted, from which one can observe the trend that, for each type of modification, higher work function, lower Li adsorption energy (stronger adsorption), showing that rising work function can serve to enhance Li adsorption on graphene. However, structural modifications also affect the coupling between Li$^+$ and the substrate, and here we find that for F-functionalized graphene, $|\Delta E_{ads}|$ is significantly smaller than $|\Delta\Phi|$, suggesting weaker coupling ($\Delta E_{cp} > 0$) according to equation (3). This is because for F-functionalized graphene, each F atom on top of the graphene plane attracts ~0.7 electrons away from the plane according to bader charge analysis[46], making the substrate more positively charged and thus weakening the electrostatic coupling.

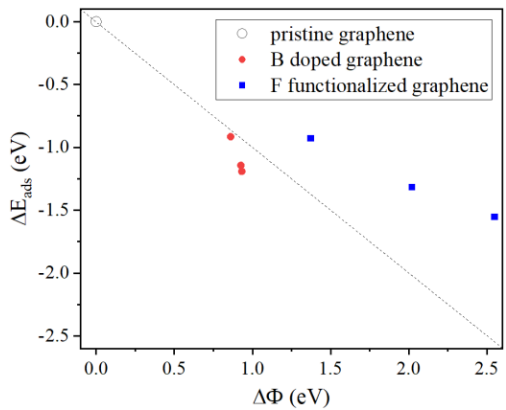

**Figure 4.** Change of Li adsorption energy versus change of work function for B-doped graphene and F-functionalized graphene.

**Limitations with our model applied to semiconductors.** Lastly, we briefly discuss the reasons why semiconductors are not included in this work. On the one hand, they are less valuable for electrode materials with lower electronic conductivity, therefore we ignore them to save computational resources. On the other hand, semiconductors will possess different Li adsorption behaviors compared with metals, which could be harder to capture. For semiconductors, if the Li 2s electron still transfers to the highest unoccupied band, the work function ($\Phi$) term in equation 3 should be replaced by electron affinity (EA) as:

$$E_{ads} = IP - EA + E_{cp} \ (8).$$

However, there are a number of concerns about using equation (8) to screen minimum Li adsorption energy for 2D semiconductors: 1) EA is related to the position of the conduction band minimum (CBM), which is notoriously hard to accurately predict using inexpensive (i.e., PBE) functionals[52], 2) mid-gap states might appear when adsorbed with Li[15], shifting EA from CBM into mid-gap states, and in turn making it no longer reasonable to screen minimum Li adsorption energy without explicitly introducing Li atom into the system, 3) Even if the effect of mid-gap states is considered, Stavric *et al.*[15] reported a line with slope around -0.92 by studying 15 2D semiconductors, and it is unclear whether the non "-1" slope is a result of data bias or other undiscovered adsorption mechanisms, and 4) In addition to charge transfer, for coupling, 2D semiconductors cannot be viewed as "conducting plates", further deviating from the image-charge coupling[41] compared with 2D metals.

# CONCLUSIONS

In summary, in order to screen Li interaction with 2D metallic materials, we have designed and tested a high-throughput screening scheme that incorporates the process of learning physics into the screening circle. First, density functional theory (DFT) and graph convolutional networks (GCN) are employed to calculate the minimum Li adsorption energies for a small set of 2D metals. Then we propose a three-step adsorption mechanism based on previous understandings of charge-transfer and ionization-coupling to explain the found linear relation between minimum Li adsorption energy and work function of 2D metals. We propose that, during adsorption, a Li atom first ionizes to $Li^+$ and an electron with the energy cost equal to the ionization potential of Li, then the electron transfers to the 2D metal and releases energy equal to the work function, and finally the $Li^+$ couples with the negatively charged 2D metal with the energy change of the coupling energy, and we use chemisorption theory to support the proposed charge transfer direction. We use a random forest model to predict and understand the coupling process, and find that variances of elemental properties of component elements and packing density are the most correlated features related to coupling. Finally, we apply our models on discovering potential high-voltage materials and find that some fluorides and chromium oxides have minimum Li adsorption energies lower than -7eV, which breaks the record, and we use the found physics as the design principle to enhance the Li adsorption ability of graphene. We show that our physics-driven models have strong ability of extrapolation and higher accuracy and transferability than purely data-driven models. We hope this work can not only deepen our understanding of the nature of Li binding and promote the application of 2D materials in Li-ion batteries, but also inspire the use of physics to simplify learning problems by decoupling the target property into simpler properties in high-throughput screening for problems that are typically challenging for conventional DFT calculations and purely data-driven machine learning.

## DATA AVAILABILITY

The training set, high-throughput screening result, GCN potential and structures of the found high-voltage materials with adsorbed Li are provided at the link:

https://github.com/shenggong1996/Screening-Li-adsorption-on-2D-metals

## AUTHOR INFORMATION

**Corresponding Author**

Professor Jeffrey C. Grossman

jcg@mit.edu

**Notes**

The authors declare no competing financial interests.

## ACKNOWLEDGMENTS

This work was supported by Toyota Research Institute. Computational support was provided by the DOE Office of Science User Facility supported by the Office of Science of the U.S. Department of Energy under Contract No. DE-AC02-05CH11231, and the Extreme Science and Engineering Discovery Environment, supported by National Science Foundation grant number ACI-1053575.